\documentclass[twocolumn,superscriptaddress]{revtex4-2}

\usepackage{amsmath}
\usepackage{amssymb}
\usepackage{graphicx}
\usepackage{color}
\usepackage{hyperref}
\usepackage{xcolor}
\usepackage{siunitx}
\usepackage[english]{babel}

\hypersetup{
    colorlinks=true,
    linkcolor=black,
    filecolor=black,
    citecolor=black,
    urlcolor=black
}

\providecommand{\ignore}[1]{}

\newif\ifcmnt
\cmnttrue
\ifdefined\cmntsoff\cmntfalse\fi

\ifcmnt
    \providecommand{\aucmnt}[1]{#1}
\else
    \providecommand{\aucmnt}[1]{}
\fi

\DeclareSIUnit\fps{FPS}
\makeatletter
\renewcommand{\p@subsection}{}
\makeatother

\newcommand{\beatNoteFreq}{\SI{49.967(85)}{\mega\hertz}}
\newcommand{\chirpDefocusThetaDeg}{\SI{-0.1665}{\degree}}
\newcommand{\chirpDefocusWxUm}{\SI{15.01}{\micro\metre}}
\newcommand{\chirpDefocusWyUm}{\SI{9.333}{\micro\metre}}
\newcommand{\chirpThetaDeg}{\SI{-0.2339}{\degree}}
\newcommand{\chirpWxUm}{\SI{15.18}{\micro\metre}}
\newcommand{\chirpWyUm}{\SI{9.049}{\micro\metre}}
\newcommand{\ditherRMSError}{\SI{0.0836(75)}{\percent}}
\newcommand{\echelleDispersionDxUmPerGHz}{\SI{4.724}{\micro\metre\per\giga\hertz}}
\newcommand{\echelleDispersionDyUmPerGHz}{\SI{0.0055}{\micro\metre\per\giga\hertz}}
\newcommand{\echelleFSRTHz}{\SI{13}{\tera\hertz}}
\newcommand{\echelleResolutionGHzPerWaist}{\SI{3.43}{\giga\hertz}}
\newcommand{\echelleWaistsCLBand}{2918{}}
\newcommand{\echelleWaistsInEOM}{11{}}
\newcommand{\echelleWaistsInTile}{83{}}
\newcommand{\harvardThetaDeg}{\SI{-1.161}{\degree}}
\newcommand{\harvardWxUm}{\SI{16.74}{\micro\metre}}
\newcommand{\harvardWyUm}{\SI{8.712}{\micro\metre}}
\newcommand{\interpDefocusThetaDeg}{\SI{-0.2615}{\degree}}
\newcommand{\interpDefocusWxUm}{\SI{15.05}{\micro\metre}}
\newcommand{\interpDefocusWyUm}{\SI{9.436}{\micro\metre}}
\newcommand{\interpThetaDeg}{\SI{-0.2213}{\degree}}
\newcommand{\interpWxUm}{\SI{15.16}{\micro\metre}}
\newcommand{\interpWyUm}{\SI{9.013}{\micro\metre}}
\newcommand{\kagomeThetaDeg}{\SI{-1.577}{\degree}}
\newcommand{\kagomeWxUm}{\SI{16.69}{\micro\metre}}
\newcommand{\kagomeWyUm}{\SI{8.587}{\micro\metre}}
\newcommand{\megabraidThetaDeg}{\SI{-0.1063}{\degree}}
\newcommand{\megabraidWxUm}{\SI{15.31}{\micro\metre}}
\newcommand{\megabraidWyUm}{\SI{8.596}{\micro\metre}}
\newcommand{\modulationFallTimeRounded}{\SI{13.0}{\nano\second}}
\newcommand{\modulationFrameRateRounded}{\SI{84}{\mega\fps}}
\newcommand{\modulationRiseTimeRounded}{\SI{11.8}{\nano\second}}
\newcommand{\phaseJumpPeakRatio}{1.604{}}
\newcommand{\phaseJumpTroughRatio}{1.562{}}
\newcommand{\psfThetaDeg}{\SI{-0.6093}{\degree}}
\newcommand{\psfWaistMajorUm}{\SI{16.11}{\micro\metre}}
\newcommand{\psfWaistMinorUm}{\SI{8.698}{\micro\metre}}
\newcommand{\tileSpanGHz}{\SI{286.9}{\giga\hertz}}
\newcommand{\tileThetaDeg}{\SI{-0.7669}{\degree}}
\newcommand{\tileWx}{\SI{16.19}{\micro\metre}}
\newcommand{\tileWy}{\SI{8.712}{\micro\metre}}
\newcommand{\toneSpacingWaists}{1.161{}}
\newcommand{\vipaDispersionDxUmPerGHz}{\SI{0.5614}{\micro\metre\per\giga\hertz}}
\newcommand{\vipaDispersionDyUmPerGHz}{\SI{202.3}{\micro\metre\per\giga\hertz}}
\newcommand{\vipaEffRoundTrips}{33{}}
\newcommand{\vipaFSRGHz}{\SI{2.26}{\giga\hertz}}
\newcommand{\vipaResolutionMHzPerWaist}{\SI{43.07}{\mega\hertz}}
\newcommand{\vipaWaistsPerFSR}{52{}}

\newcommand{\scaledVipaEffRoundTrips}{1000{}}
\newcommand{\scaledVipaFSRGHz}{\SI{9.99}{\giga\hertz}}
\newcommand{\scaledVipaPredRiseNs}{\SI{75}{\nano\second}}
\newcommand{\scaledVipaResolutionMHzPerFWHM}{\SI{7.49}{\mega\hertz}}
\newcommand{\scaledModulationFrameRateRounded}{\SI{13}{\mega\fps}}

\begin{document}

\title{Ultrafast and high resolution spatial light modulation for cold atoms}

\newcommand{\harvard}{Department of Physics, Harvard University, Cambridge, MA, USA}
\newcommand{\harvardqse}{Quantum Science and Engineering, Harvard University, Cambridge, MA, USA}

\author{Alexander Dennisovich Deters}
\affiliation{\harvard}
\affiliation{\harvardqse}

\author{Yanfei Li}
\affiliation{\harvard}

\author{Alexander Douglas}
\affiliation{\harvard}

\author{Markus Greiner}
\email{mgreiner@g.harvard.edu}
\affiliation{\harvard}

\author{Aaron W. Young}
\email{aaron\_young@g.harvard.edu}
\affiliation{\harvard}

\date{\today}

\begin{abstract}

    Programmable arrays of ultracold atoms are a leading platform for quantum computation and simulation, enabling state-of-the-art implementations of quantum error correction~\cite{bluvsteinFaulttolerantNeutralatomArchitecture2026, reichardtFaulttolerantQuantumComputation2025}, and analog simulations of Hubbard models that address open problems in condensed matter physics~\cite{grossQuantumSimulationsUltracold2017, kendrickPseudogapFermiHubbard2026}.
    In these systems, all local control is mediated through precisely shaped optical fields, and so the challenge of managing many-body quantum states becomes an exercise in optical design.
    In particular, one wishes for fast, flexible control with low disorder and heating, and access to large arrays with many atoms.
    An ideal optical system therefore must generate arbitrary patterns with high spatial resolution and low disorder, and alter these patterns on a timescale that is faster than the relevant atomic dynamics.
    Here, we present an optical system that is comparable to previous approaches in scale, while advancing all other axes.
    We demonstrate arbitrary pattern generation with $10^{-3}$ intensity resolution, a frame rate of $>\modulationFrameRateRounded$ (megaframes per second), and a spatial resolution of $\echelleWaistsInTile\times\vipaWaistsPerFSR$ beam waists (with $\echelleWaistsInEOM \times \vipaWaistsPerFSR$ waists accessible via a single \SI{40}{\giga\hertz} electro-optic modulator).
    These capabilities unlock a new class of experiments.
    We develop and numerically validate a scheme for fully programmable Hubbard models, with time-dependent control over local chemical potentials, tunneling amplitudes, on-site interactions, and patterns of artificial magnetic flux.
    The same architecture performs fast, arbitrary permutations of tweezers in 2D, decoupling optical constraints from the design of high-rate error-correcting codes.
\end{abstract}

\maketitle

\section{Introduction}

Optical modulators that fulfill the typically conflicting demands of arbitrary pattern generation, high spatial and intensity resolution, and rapid refresh rates unlock transformative capabilities for programmable atom arrays and optical lattices.
In analog simulation of Hubbard models, the projection of arbitrary, time-varying potentials with low disorder enables improved state preparation~\cite{cotlerQuantumVirtualCooling2019, langbehnDiluteMeasurementInducedCooling2024, kamalFloquetFluxAttachment2024, palmGrowingExtendedLaughlin2024, defossezDynamicRealizationMajorana2025} and provides access to arbitrary observables~\cite{tranMeasuringArbitraryPhysical2023, markEfficientlyMeasuringdWave2025}. These tools also significantly expand the accessible model space to include more lattice geometries~\cite{xuFrustrationDopinginducedMagnetism2023, joUltracoldAtomsTunable2012a}, including multi-band systems~\cite{macridinPhysicsCupratesTwoband2005, wirthEvidenceOrbitalSuperfluidity2011, lebratFerrimagnetismUltracoldFermions2026}, and the non-periodic structures required to directly emulate localized lattice defects \cite{ weiKondoImpurityAttractive2025,amaricciEngineeringKondoImpurity2025} and certain molecules \cite{arguello-luengoAnalogueQuantumChemistry2019, luhmannEmulatingMolecularOrbitals2015, maskaraProgrammableSimulationsMolecules2025} [Fig.~\ref{fig:setup}(b)].
In digital quantum computation, rapid, unconstrained two-dimensional routing of isolated tweezers is essential for efficient fault-tolerant logic.
Specifically, this freedom enables the realization of asymptotically good quantum low-density parity-check (qLDPC) codes~\cite{leverrierQuantumTannerCodes2022, dinurGoodQuantumLDPC2023}, surpassing hypergraph or lifted product codes that are compatible with crossed acousto-optic deflectors~\cite{xuConstantoverheadFaulttolerantQuantum2024}.

However, current optical modulators that permit arbitrary intensity patterns are limited in speed, with digital micromirror devices (DMDs) and liquid crystal on silicon (LCOS) spatial light modulators (SLMs) reaching frame rates of up to $\sim50$~kHz and $\sim1$~kHz, respectively~\cite{parkTechnologiesModulationVisible2024, brandtSpatialLightModulators2011, knottnerusParallelAssemblyNeutral2025, linAIEnabledParallelAssembly2025}.
Acousto-optic deflectors (AODs) can achieve faster modulation~\cite{Heberle2016ElectroopticAA, luAstigmatismfree3DOptical2026, guoAcoustoopticLens3D2025, picardThreedimensionalAcoustoopticDeflector2026}, but are limited in the patterns they can produce in 2D by the outer product structure imposed by crossed AODs~\cite{bluvsteinQuantumProcessorBased2022, xuConstantoverheadFaulttolerantQuantum2024}.
Additionally, high resolution AODs can only achieve frame rates on the order of $\sim100$~kHz, which is limiting in some tweezer array systems~\cite{Heberle2016ElectroopticAA, luAstigmatismfree3DOptical2026, guoAcoustoopticLens3D2025, picardThreedimensionalAcoustoopticDeflector2026, yanTwoDimensionalProgrammableTweezer2022}.
For large arrays of tweezers, all of the above approaches are typically limited to an RMS inhomogeneity on the scale of $\gtrsim 1\%$ in intensity~\cite{manetschTweezerArray61002025, bluvsteinLogicalQuantumProcessor2024, chewUltrapreciseHolographicOptical2024}.

Here, we present an alternative approach based on high speed frequency modulation and diffractive optics, which we will refer to as a dispersive spatial light modulator (dSLM).
State-of-the-art telecom modulators routinely span the C- and L-bands, exceeding $10$~THz, with fast, high-resolution amplitude and phase control \cite{winzerFiberopticTransmissionNetworking2018}.
Standard fiber amplifiers can boost these signals to high powers ($>10$ W) across the entire bandwidth.
When paired with an appropriate dispersive element that maps frequency to position in 2D, this yields a fast, high-resolution, and high-brightness display.
The approach of using frequency modulation for spatial light control was recently demonstrated in a re-imaging phased array (RIPA), achieving arbitrary pattern generation using an $8\times9$ array of phase coherent emitters with a refresh time of $t_{\text{rise}} = 44(1)$~ns, implying a frame rate of $t_{\text{rise}}^{-1} = 22.7(5)$~MFPS \cite{wei10MegahertzSpatial2026}.
In this work, we pursue this paradigm with a distinct architecture which converts a large bandwidth to high spatial resolution, while realizing a frame rate of $>\modulationFrameRateRounded$.

\section{The dispersive spatial light modulator}

\begin{figure}[htbp]
    \centering
    \includegraphics[width=89mm]{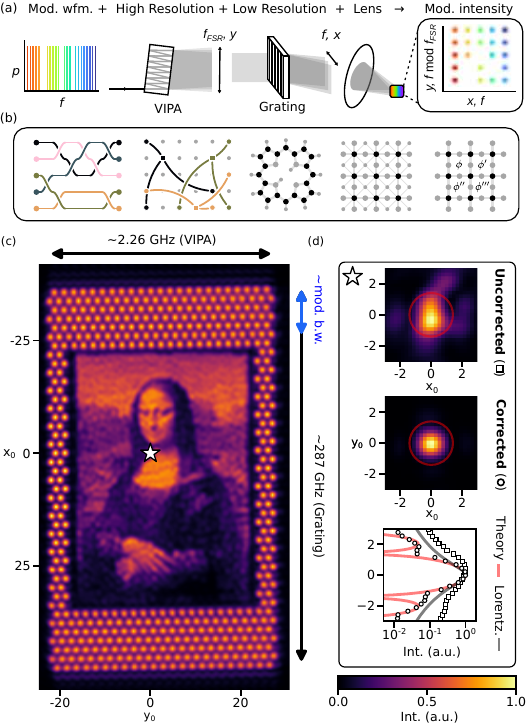}
    \caption{Principles and optical performance of the dSLM. (a) A broadband modulated waveform is split in the $x$ ($y$) directions by dispersive elements with low (high) resolution and large (small) FSR.
        (b) The resulting display can produce arbitrary, time-varying potentials that are of broad use to atom arrays and optical lattices, including, from left to right: arbitrary permutations of tweezers; qLDPC codes unconstrained by AODs; non-repeating potentials of relevance to simulating molecules and lattice defects; the realization of lattices with multiple bands; Floquet modulation for arbitrary patterns of flux.
        (c)
        To characterize the optical performance of the system beyond the bandwidth of the single modulator used in this work (marked in blue), we scan a single spot across the full field of view of the system to produce an image (Appendix~\ref{sec:postprocessing}).
        Axes displayed in units of beam waists $x_0 = x / w_{x,0}$,  $y_0 = y / w_{y,0}$.
	(d) Representative PSF [at the location marked by a star in (c)] before and after SLM correction near the reference wavelength. Red circle indicates first null of the diffraction-limited PSF along the VIPA ($y$) axis. Linecuts display the measured corrected (circles) and uncorrected (squares) PSFs, theoretical diffraction-limited Airy profile (red line), and a Lorentzian profile with equal FWHM (gray line).
        Colorbars are shared across this figure.
    }
    \label{fig:setup}
\end{figure}

Our system pairs a virtually imaged phased array (VIPA)~\cite{shirasakiLargeAngularDispersion1996, shirasakiVirtuallyImagedPhased1999} with a diffraction grating, and leverages both the high frequency resolution of the VIPA and the high bandwidth of the grating.
The VIPA operates by repeated reflection of a beam between two surfaces, where each bounce produces an additional copy of the beam with a well defined phase offset which varies with laser frequency~\cite{shirasakiLargeAngularDispersion1996}.
The result is a frequency-to-angle mapping (in the $y$ axis) that repeats every free spectral range (FSR), when the phase offset between adjacent emitters changes by $2\pi$.
Subsequently, a diffraction grating deflects the spot in an orthogonal direction (the $x$ axis), with a frequency resolution that is matched to the FSR of the VIPA.
Focusing the output of this system with a lens produces a unique mapping between frequency and position in two dimensions [Fig.~\ref{fig:setup}(a)]. 
A similar mapping has been used for spectroscopy~\cite{diddams_molecular_2007, leung2025viper, sadiek_air-spaced_2024} and nonmechanical LiDAR~\cite{okanoSweptSourceLidar2020,
liSolidstateFMCWLiDAR2021, dostartSerpentineOpticalPhased2020}.
In this work, the VIPA deflects the spot by one waist for a frequency shift of $\vipaResolutionMHzPerWaist$, and has an FSR of $\simeq\vipaFSRGHz$.
The diffraction grating deflects the spot by one waist every $\echelleResolutionGHzPerWaist$, and has an FSR of $\simeq \echelleFSRTHz$.
Together, the optical elements produce a display with $\simeq\echelleWaistsCLBand \times \vipaWaistsPerFSR$ resolvable spots over a bandwidth equivalent to the C+L band, with $\echelleWaistsInTile \times \vipaWaistsPerFSR$ within the field of view of the off-the-shelf imaging components used in this work.

Care must be taken to produce a diffraction-limited spot using the dSLM.
Note that a uniformly reflective coating yields an exponential decay in the intensity of the emitters, yielding a Lorentzian beam profile with slowly decaying tails.
We instead use a gradient coating \cite{shirasakiVirtuallyImagedPhased1999} to achieve an approximately uniform illumination across the emitters (Appendix~\ref{sec:layout}).
This yields the same sinc profile as any uniformly filled aperture, and may be apodized to a Gaussian.
Additionally, the multiple round trips through the VIPA result in extreme sensitivity to surface polishing, and thus significant wavefront errors.
Crucially, these errors are only weakly dependent on frequency, and so a static correction applied with an LCOS SLM allows us to recover diffraction-limited performance across the entire bandwidth of the system [Figs.~\ref{fig:setup}(c) and \ref{fig:setup}(d)].

\section{Ultrafast spatiotemporal control}

\begin{figure*}[htbp]
    \centering
    \includegraphics[width=180mm]{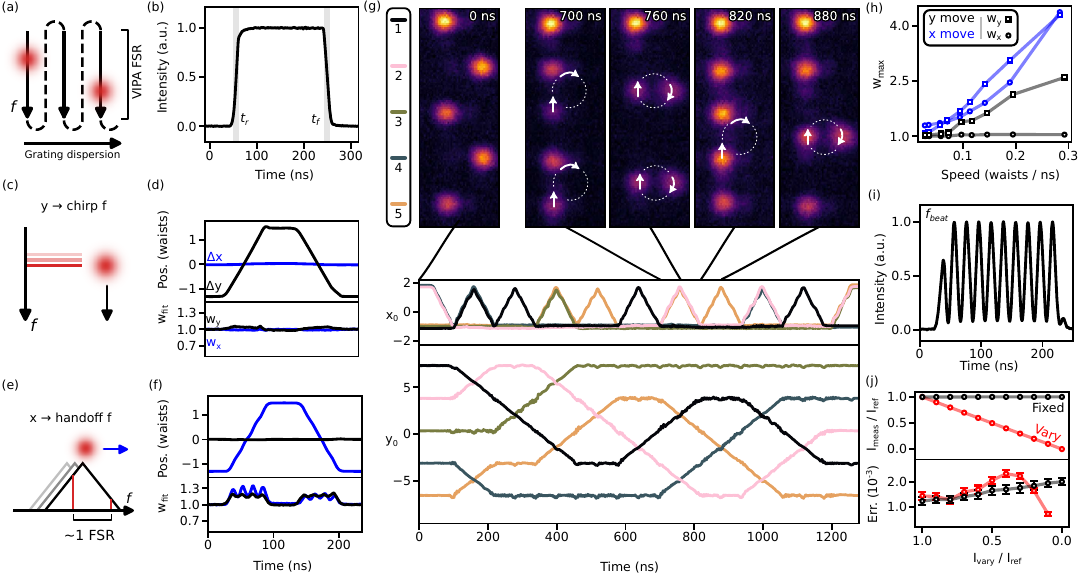}
    \caption{Ultrafast spatiotemporal control of tweezers. (a) Individual tones are dispersed periodically in the $y$ direction by the VIPA and continuously in the $x$ direction by the grating with increasing frequency. This creates a unique mapping of frequency to position.
        (b) Intensity near the center of a single tweezer as it is turned on for $200$~ns and then extinguished. Tweezers are generated in real-time with a \SI{40}{\giga\hertz} Mach-Zehnder modulator. The resulting 10-90\% rise time of $t_{\mathrm{rise}}=\modulationRiseTimeRounded$ and fall time of $t_{\mathrm{fall}}=\modulationFallTimeRounded$ correspond to an effective frame rate of $t_{\mathrm{rise}}^{-1} > \modulationFrameRateRounded$.
        (c) Chirping a frequency within the VIPA FSR continuously moves a tweezer along the $y$ direction.
        (d) Demonstrating a move in the VIPA ($y$) direction, motion recorded with a fast single-pixel camera (Appendix~\ref{sec:singlepixel}). Upper plot shows tweezer displacement with mean value subtracted $(\Delta x, \Delta y)$ and lower plot contains the beam waist normalized by its initial value $(w_x, w_y)$.
        (e) Smoothly handing off between tones separated by the VIPA FSR continuously moves a tweezer along the $x$ direction.
        (f) Demonstrating a move in the grating ($x$) direction. A small frequency correction is included to compensate for residual vertical dispersion of the grating (Appendix~\ref{sec:handoff}).
        (g) Simultaneous, arbitrary braiding of $5$ tweezers on a $\SI{1.2}{\micro\second}$ timescale. Upper plot displays selected frames and the lower plot displays fitted beam positions in units of the beam waist.  
        (h) Peak normalized beam waist ($w_\text{max}$) as a function of movement speed. Movement along the $y$ axis does not significantly increase the waist size in the $x$ axis whereas moving in the $x$ axis affects both axes. $x$ movement requires a handoff between discretely spaced points, and so incurs a small constant factor increase in $x$ waist even at low speeds.
    (i) Interference beat note between two tweezers spaced by $50$ MHz, with a fitted value of $f_{\mathrm{beat}} = \beatNoteFreq$.  (j) We calibrate the intensity vs RF drive amplitude and hold two tones fixed while varying a third. The mean normalized RMS error is $\ditherRMSError$ when homogenizing three tones. Videos for the above measurements are attached as supplemental media.}
    \label{fig:performance}
\end{figure*}

Arbitrary intensity distributions are produced by appropriately shaping the power spectral density of the light [Fig.~\ref{fig:setup}(a)], in this case with a high frequency fiber-based Mach-Zehnder modulator (MZM).
Note that in this proof-of-principle demonstration, we use a single MZM with a modulation bandwidth of $40$~GHz, which addresses a region spanning $\echelleWaistsInEOM \times  \vipaWaistsPerFSR$ beam waists.
In order to test the optical performance of the system, we stitch multiple regions together by scanning the center wavelength of the laser [Figs.~\ref{fig:setup}(c) and \ref{fig:hubbard}(a)].
In the future, multiple modulators can readily be combined with standard wavelength division multiplexers to span a larger region~\cite{winzerFiberopticTransmissionNetworking2018}.

The response of the system is dominated by the propagation time through the VIPA, corresponding to $\simeq \vipaEffRoundTrips $ repeated $2\times 4.5$~cm-long round trips (equal to the resolvable spots in the $y$ direction times $2 / \pi$).
To characterize the performance of the dSLM in the time domain, we use a fast, single-pixel camera to record videos faster than this timescale (Appendix~\ref{sec:singlepixel}).
When the amplitude or frequency of the drive beam is changed, emitters in the phased array are effectively updated sequentially upon each round trip of the beam.
Turning the beam on therefore yields a quadratic response in intensity, because the number of interfering emitters grows linearly in time.
We measure a 10-90\% rise time of $\modulationRiseTimeRounded$ and fall time of $\modulationFallTimeRounded$ when turning the beam on or off, which corresponds to a frame rate of $>\modulationFrameRateRounded$, in line with expectations [Fig.~\ref{fig:performance}(b)]. 

Even faster features can be generated by leveraging the fact that an update sweeps linearly across the VIPA.
For example, one can halve the response time of the VIPA by jumping the phase by $\pi$, resulting in destructive interference between the first and second half of the phased array when the phase slip has traversed half of the VIPA (Appendix~\ref{sec:phasejump}).
In a similar setup, this behavior could be used to generate patterns in 3D by leveraging the fact that a sufficiently fast chirp in the drive frequency yields a focal shift, similar to AODs~\cite{luAstigmatismfree3DOptical2026, guoAcoustoopticLens3D2025, picardThreedimensionalAcoustoopticDeflector2026}.

The dSLM can display a video by cycling through different intensity patterns with a frame rate set by the above rise time. 
Coherent transport of atoms benefits from continuous motion, and the ability to generate spots closer than is resolvable.
The dSLM achieves this in distinct ways along the two axes.
In the VIPA axis, the realizable positions are continuous, and ramping the frequency (chirping) creates continuous motion without sacrificing the effective resolution of the device.
Along the grating axis, the realizable positions are discrete. However, by choosing the VIPA FSR to be smaller than the grating resolution, one can create continuous motion by handing off between tones spaced by one FSR of the VIPA [Figs.~\ref{fig:performance}(c)--\ref{fig:performance}(f) and Appendix~\ref{sec:handoff}].
In this scheme, $x$ moves incur a small constant factor increase in the effective beam waist [Fig.~\ref{fig:performance}(h) and Appendix~\ref{sec:handoff}]; otherwise, the resolution of the display remains approximately unchanged as a function of speed for moves as fast as 100 waists$/\mu$s. 
For faster moves, the effective aperture of the VIPA is reduced, reducing the resolution of the display along the $y$ axis [Fig.~\ref{fig:performance}(h)].

In Fig.~\ref{fig:performance}(g), we show that the dSLM is capable of performing a series of arbitrary permutations of 5 spots arranged in 2D in $\sim \SI{1}{\micro\second}$.
Note that, in contrast to crossed AOD arrays~\cite{xuConstantoverheadFaulttolerantQuantum2024, constantinidesOptimalRoutingProtocols2024}, a single arbitrary permutation in 2D using constant velocity moves can be performed in sublinear time with respect to system size, as is important for certain efficient implementations of error correction and fault-tolerant logic~\cite{leverrierQuantumTannerCodes2022, dinurGoodQuantumLDPC2023}.

A critical advantage of the dSLM architecture is that the frequency separation between adjacent spots can be made to be large compared to all relevant atomic dynamics while maintaining high spatial resolution.
For example, two spots separated by only $\toneSpacingWaists$ waists in the $y$ axis produce a beat note at $\SI{50}{\mega\hertz}$ in our system [Fig.~\ref{fig:performance}(i)].
Note that the minimum separation in the $x$ axis corresponds to $\ge1$~FSR, and is beyond the bandwidth of our measurement apparatus.
Nearby spots are therefore effectively non-interfering when averaging over the $\rm\gtrsim1~\mu s$ timescales relevant for atomic motion, allowing for fine-tuned control over relative intensity.
In Fig.~\ref{fig:performance}(j), we demonstrate intensity homogenization over 3 sites at the $10^{-3}$ level.
Similar control can be realized for larger arrays and in 2D, provided one has a sufficiently high resolution microwave source to drive the MZM.

\section{Programmable Hubbard models}

\begin{figure*}[!htb]
	\centering
	\includegraphics[width=180mm]{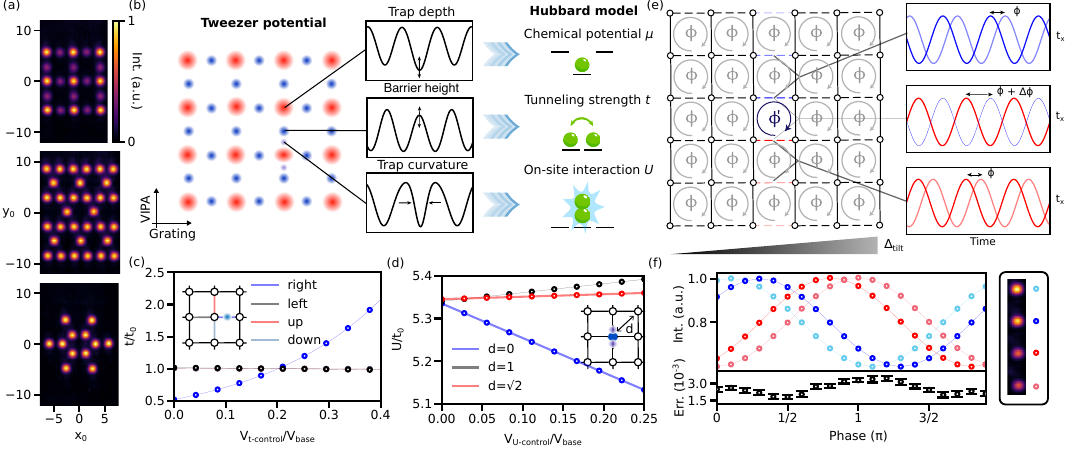}
    \caption{Programmable potentials for analog quantum simulation of the Hubbard model. (a) Example lattice geometries projected by the dSLM; individual FSRs are modulated separately and stitched together with carrier and sidebands cropped. Top to bottom: Lieb (programmable square), kagome, and molecular-orbital lattices.
		(b) Scheme for a fully programmable square lattice Hubbard model using tweezer arrays.
        The local chemical potential is controlled by a base set of tweezers (red) that define the vertices in the lattice, tunneling amplitudes by interlaced $t$-control tweezers (blue) that tune the height of a tunnel barrier, and on-site interactions by $U$-control tweezers (purple) that modify the trap curvature.
        Numerical simulations show that all of the above parameters can be tuned independently [(c) and (d)].
		(c) Single-link tunneling control. A $t$-control tweezer placed $1w_0$ away from the base tweezers independently tunes the right-link tunneling amplitude (blue) relative to the background tunneling $t_0$ (red, black, cyan).
		(d) Single-site interaction control. $U$-control tweezers placed $0.6w_0$ above and below the central site tune the central interaction strength (blue) relative to neighboring sites (red, black).
        (e) Floquet engineering of programmable artificial magnetic flux. A linear gradient $\Delta_{\mathrm{tilt}}$ is applied along $x$, and $t_x$ is resonantly modulated. Link-dependent phase offsets along $y$ generate an Aharonov--Bohm phase associated with each plaquette; judicious choices of the phases yield a modified central flux of $\phi'$ while maintaining a background flux of $\phi$.
    (f) Experimental modulation of four tweezer spots with spatially dependent phases. Top to bottom, spots $1,2$ and $3,4$ have a phase offset of $\Delta\phi=\pi/5$, while spots $2,3$ have $\Delta\phi'=\pi/2$, realizing a minimal flux-defect pattern. Error is the mean absolute deviation of the measured amplitudes from their targets, averaged over the four spots and $20$ repetitions; error bars denote the standard error of the mean.}
	\label{fig:hubbard}
\end{figure*}

\begin{figure*}[!htb]
    \centering
    \includegraphics[width=180mm]{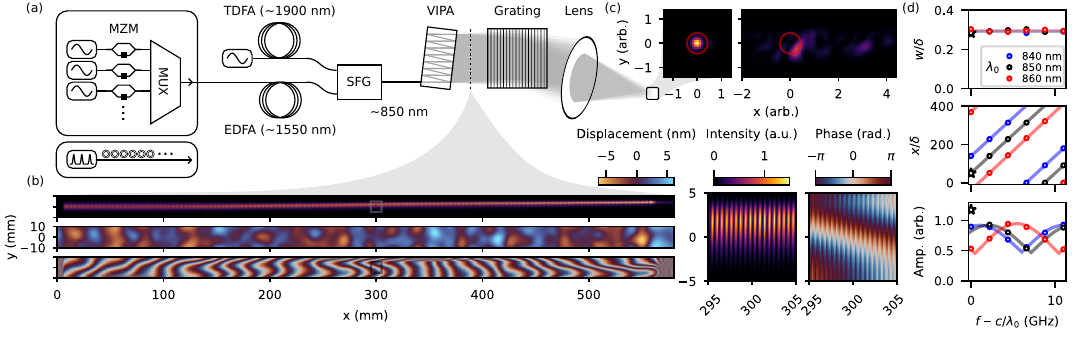}
    \caption{Scaling to megapixel resolution. (a) Standard telecom techniques provide modulation over $>10$~THz of bandwidth, either via multiplexed arrays of Mach-Zehnder modulators, or architectures involving frequency combs coupled to large arrays of ring resonators. This can be paired with high power fiber amplifiers, efficient, high bandwidth frequency conversion, and the appropriate diffractive optics, to achieve high power and high resolution modulation at convenient wavelengths for cold atom experiments. Here, we consider $\sim850$~nm for trapping rubidium via sum frequency generation (SFG) using thulium and erbium doped fiber amplifiers (TDFA and EDFA). (b) Numerical simulations of a physically realistic VIPA with imperfect surface polishing (middle) yield significant wavefront error (bottom and right inset), and some redistribution of intensity at the output facet (top and left inset). (c) These errors produce an aberrated spot (right). However, diffraction-limited performance can be recovered by applying a correction with an  SLM (left, intensity divided by 8 in comparison to the uncorrected case). Red circle indicates the first null of the diffraction-limited Airy disk. (d) Critically, this correction is effective over large bandwidths. For a correction that is optimized at a reference wavelength of $\lambda_{0} = 850$~nm (starred points), there are negligible changes in waist ($w$), frequency ($f$) to position ($x$) conversion, and amplitude (amp.) at a given frequency over a range spanning $\pm10$~nm.
    This means that the same static correction can be used over $> 10$ THz of bandwidth. Lines indicate mean value in the upper plot, the result of a sawtooth fit with period set by the FSR of the VIPA in the middle plot, and a shape factor corresponding to the highest peak when scanning a comb spaced by the VIPA FSR across a Bessel function in the lower plot; $\delta$ is an arbitrary length unit. The system presented in these simulations enables a resolution of $>1000\times1000$ beam waists with a rise time of $\scaledVipaPredRiseNs$. Colorbars are shared across this figure.}
    \label{fig:scaling}
\end{figure*}

The ability to generate arbitrary patterns with low disorder is particularly useful in the context of analog quantum simulation of the Hubbard model and its variants, where these patterns allow one to realize different lattice or defect models.
This approach has been explored using crossed AODs in the past, where the limitations in homogeneity were overcome by rastering a 1D array~\cite{yanTwoDimensionalProgrammableTweezer2022}.
However, the smaller separation between the response time of the optical system and the relevant atomic dynamics in that setting limited system size.
The ability to rapidly and flexibly modulate these patterns using the dSLM further enables the study of time-dependent phenomena, including Floquet engineering and quench dynamics.

As an example, in Fig.~\ref{fig:hubbard}(b) we present and numerically simulate a scheme for nearly arbitrary, time-dependent control over all parameters in a square lattice Hubbard model using a dSLM (Appendices \ref{sec:mlwf}--\ref{sec:local_flux}).
Specifically, the model is one in which atoms occupy vertices $i$ on a graph (in this case a square lattice), can hop between vertices connected by an edge $\{ij\}$, and experience a local interaction when multiple atoms occupy the same site.
Each vertex can have a unique local potential $\mu_i$ and interaction strength $U_i$, and each edge is associated with a tunneling strength $t_{ij}$.
By introducing more optical spots than vertices, we can tune $\mu_i$, $U_i$, and $t_{ij}$ independently.
$\mu_i$ is primarily tuned via the intensity of a tweezer that defines a given lattice site, $U_i$ via secondary spots that tune the curvature of the site, and $t_{ij}$ via the height of a barrier between sites.

In practice, the different parameters are coupled, and we numerically optimize the potential to produce a given set of parameters [Figs.~\ref{fig:hubbard}(c) and \ref{fig:hubbard}(d), Appendix \ref{sec:hubbard_opt}].
The range over which the different parameters can be tuned depends on specific details of the atomic species and optical system.

We can take advantage of time-dependent control of the above parameters to achieve fundamentally new capabilities.
For example, in Figs.~\ref{fig:hubbard}(e) and \ref{fig:hubbard}(f) we show that the global Floquet modulation techniques that have enabled simulations of lattice models with artificial gauge fields \cite{aidelsburgerMeasuringChernNumber2015, taiMicroscopyInteractingHarper2017} can be extended to locally varying gauge fields.
Specifically, by using the dSLM to independently control the driving phase of each link, one can generate nearly arbitrary patterns of flux. 
This represents a significant expansion of the kinds of dynamics that can be simulated, including intriguing scenarios where one can pin and manipulate fractional excitations, for example in a fractional Chern insulator state \cite{wangMeasurableSignaturesBosonic2022}, as would be required for braiding anyons \cite{nayakNonAbelianAnyonsTopological2008, kimProgrammableLatticesNonAbelian2026}.
More broadly, local, time-dependent control of a Hubbard model opens the door to a wealth of directions involving new schemes for state preparation~\cite{cotlerQuantumVirtualCooling2019, langbehnDiluteMeasurementInducedCooling2024, kamalFloquetFluxAttachment2024, palmGrowingExtendedLaughlin2024, defossezDynamicRealizationMajorana2025}, driven defects~\cite{hubnerFloquetengineeredPairSingleparticle2022}, and quench-based measurements of exotic observables~\cite{caioTopologicalMarkerCurrents2019, tranMeasuringArbitraryPhysical2023, umucalilarBulkDensitySignatures2023, unalCircularDichroismEdge2025, markEfficientlyMeasuringdWave2025}.

\section{Scaling to megapixel resolution}

The dSLM architecture is readily scalable, including at convenient wavelengths for atom trapping, by taking advantage of broadband telecom modulators, high power fiber-based amplifiers, broadband frequency conversion, and improved optical polishing.
Standard telecom modulators can span the entire C- and L-bands with amplitude modulated signals, and erbium doped fiber amplifiers (EDFAs) can boost these signals to high power ($>10~$W)~\cite{winzerFiberopticTransmissionNetworking2018}.
Sum frequency generation with a single pass of the telecom signal and a single-frequency, cavity enhanced pump can convert these signals to convenient wavelengths for atom trapping with high efficiency and low intermodulation.
For example, to obtain typical trapping wavelengths for rubidium atoms of around 850~nm, one could perform sum frequency generation with an L/C-band signal and a thulium doped fiber amplifier (TDFA) centered at $\sim1900~$nm.

State-of-the-art techniques for optical polishing (e.g. via ion beam figuring) can realistically produce flat surfaces with an RMS surface error of $<\lambda/100$~\cite{frostLargeAreaSmoothing2009}.
Even better performance is possible in high end applications like EUV lithography, where $\lesssim0.1~$nm RMS surface error is required~\cite{louisNanometerInterfaceMaterials2011}.
In Fig.~\ref{fig:scaling}, we simulate the performance of a hypothetical system based on the above parameters.
Specifically, we consider $n_{RT} \simeq \scaledVipaEffRoundTrips$ round trips through a large air-gapped VIPA with a thickness of $d = 15$~mm, and an edge length of $l=60~$cm, which produces a free spectral range of $\Delta_{FSR} = \scaledVipaFSRGHz$ and a FWHM frequency resolution of $\delta f = \scaledVipaResolutionMHzPerFWHM$.
We assume surface errors with a correlation length of $10~$mm, and an RMS variation of $\delta_{RMS} = 4$~nm in VIPA thickness.
Pairing this VIPA with an optical grating with $\delta f = 10~$GHz would yield a $>1$ megapixel display ($>1000\times1000$ beam waists) with a frame rate of $\scaledModulationFrameRateRounded$.

Note that the above performance relies on two key insights:
First, because diffraction in the VIPA leads to the emitters lying in a (virtual) tilted plane, this does not significantly degrade the performance of the VIPA, and one can allow for a greater number of round trips than the limit implied by the Rayleigh range of the input beam.
Second, although the above setup produces a significantly aberrated spot [Fig.~\ref{fig:scaling}(c)], these aberrations are very insensitive to frequency, and can be corrected with a static spatial light modulator or phase mask.
In Fig.~\ref{fig:scaling}(d), we show that with the same correction, the performance of the system is not significantly impacted over a range spanning from 840~nm to 860~nm, or $\sim8~$THz.
Note that while phase errors are readily correctable, phase errors with higher spatial frequency can lead to a redistribution of intensity at the VIPA output after multiple round trips. 
Although such errors can be corrected using an SLM that modulates both amplitude and phase, this comes at the cost of system efficiency.
Nevertheless, this suggests that more sophisticated optical setups could push to resolutions beyond even the 1 megapixel example presented above. 

Further scaling is possible by intermittently extracting and reimaging a beam in the VIPA, or by tiling multiple phase coherent VIPAs next to each other.
Ultimately, the VIPA provides a tradeoff between complexity and resolution.
For the fastest possible display, one could use an array of waveguide modulators, with one modulator per emitter~\cite{panuskiFullDegreeoffreedomSpatiotemporal2022, zhaoIntegratedPhotonicsPlatform2025}.
By pairing each such emitter with a VIPA, one can trade off between speed and spatial resolution by a factor corresponding to the number of round trips through the VIPA.

\section{Conclusions and outlook}

In this work, we have established a new approach to spatial light modulation involving high power and high bandwidth telecom laser sources, frequency conversion, and diffractive optics.
The result is a significant enhancement in the speed, scale, and precision with which one can generate high power optical potentials.
As a proof of principle, we demonstrate a display with a refresh time of $\modulationRiseTimeRounded$, a spatial resolution of $\echelleWaistsInTile \times  \vipaWaistsPerFSR$ beam waists, and $10^{-3}$ level intensity resolution.

These properties are particularly useful for digital and analog quantum computing with neutral atoms.
We show that one can perform arbitrary permutations of atoms in sublinear time, and develop a framework for using these devices to achieve nearly arbitrary, time-dependent control of the parameters in a Hubbard model.
We further expect the above approach to be broadly useful in applications involving fast timescales and structured light.
For example, in scanning microscopy and optical coherence tomography, the system's $>\modulationFrameRateRounded$ frame rate enables significantly enhanced data acquisition rates \cite{kleinHighspeedOCTLight2017}.

Critically, the presented architecture is compatible with much higher spatial resolution while maintaining similar speeds.
One outstanding challenge is to maintain the achieved level of homogeneity in a larger system, which requires large scale, low distortion RF control.
However, even with substantially lower homogeneity, such a system has significant utility for neutral and charged atom-based digital quantum computing, where programmable optical potentials with a fast refresh rate can be used to control atomic motion, or for local addressing via light shifting~\cite{bluvsteinQuantumProcessorBased2022, liHighRateHighFidelityModular2024, munizHighFidelityUniversalGates2025}.

\begin{acknowledgments}
We thank the Doyle, Lon\v{c}ar, and Lukin groups for sharing critical equipment for this work, particularly Matthew Bilotta, Brandon Grinkemeyer, Pavel Kurilovich, Giseok Lee, Mingda Li, Xudong Li, and Scarlett Yu. 
We thank Christopher Myatt and Precision Photonics for providing the VIPA.
We further acknowledge Alexandra Geim, Abhishek Karve, Zeyang Li, Mikhail Lukin, Nishad Maskara, Adam Shaw, Jon Simon, Vladan Vuleti\'c, Xin Wei, Muqing Xu, Hengyun Zhou, and Xiangying Zuo as well as the entire Greiner lab for helpful discussions. We acknowledge support from QuEra grant No. A57912 and the Intelligence Community Postdoctoral Research Fellowship Program at Harvard administered by Oak Ridge Institute for Science and Education (ORISE) through an interagency agreement between the U.S. Department of Energy and the Office of the Director of National Intelligence (ODNI) (A.W.Y.).
\end{acknowledgments}

\section*{Author Contributions}

M.G. and A.W.Y. conceived the project and supervised the study.
A.D.D. and A.W.Y. performed the experiments and analyzed the data.
Y.L. and A.W.Y. performed the numerical simulations.
All authors contributed to the interpretation of the results and production of the manuscript.

\section*{Competing Interests}

M.G. is co-founder, shareholder, and consultant of QuEra Computing.
A.W.Y. is currently affiliated with Farfield.
A.D.D., Y.L., A.D., M.G., and A.W.Y. are inventors on a provisional patent related to this work.

\section*{Data Availability}
The data that support the findings of this article, including figure source data and analysis scripts, will be openly available in Zenodo at publication.

\bibliography{VIPA_SLM}

\appendix

\section{Units}
In this work we define the frame rate as the inverse of the rise time. We define the spatial resolution of the system in terms of the $\frac{1}{e^2}$ radii given by an elliptical Gaussian fit of the PSF shown in Fig.~\ref{fig:setup}(d).

\section{Optical layout}
\label{sec:layout}
\begin{figure}[!htb]
    \centering
    \includegraphics[width=\linewidth]{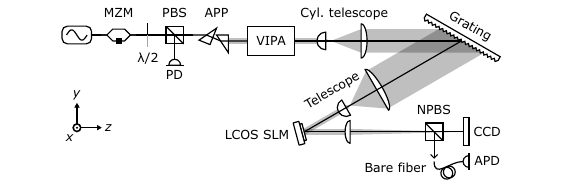}
    \caption{Full optical layout of the system, including Mach-Zehnder modulator (MZM), polarizing and non-polarizing beam splitters (PBS, NPBS), photodiode (PD), anamorphic prism pair (APP), VIPA, cylindrical telescope, grating, SLM, camera (CCD), and avalanche photodiode (APD).}
    \label{sfig:layout}
\end{figure}

The optical design of the system (Fig. \ref{sfig:layout}) is compatible with simultaneous operation over $>10~$THz of bandwidth. Here, as a proof of principle, our signal path is composed of a single Mach-Zehnder modulator with 40 GHz of bandwidth, which is paired with a $780~$nm laser source that is tunable over $>300~$GHz.

After passing through an MZM, the beam is expanded along one axis in a 4:1 anamorphic prism pair.
This ensures that the beam does not diffract significantly in the $x$ axis (the axis orthogonal to the VIPA deflection) as it traverses the VIPA.
Note that diffraction in the $y$ axis (along which the VIPA deflects) does not substantially modify the resolution of the VIPA, as discussed in the main text.

\begin{figure}[!htb]
    \centering
    \includegraphics[width=\linewidth]{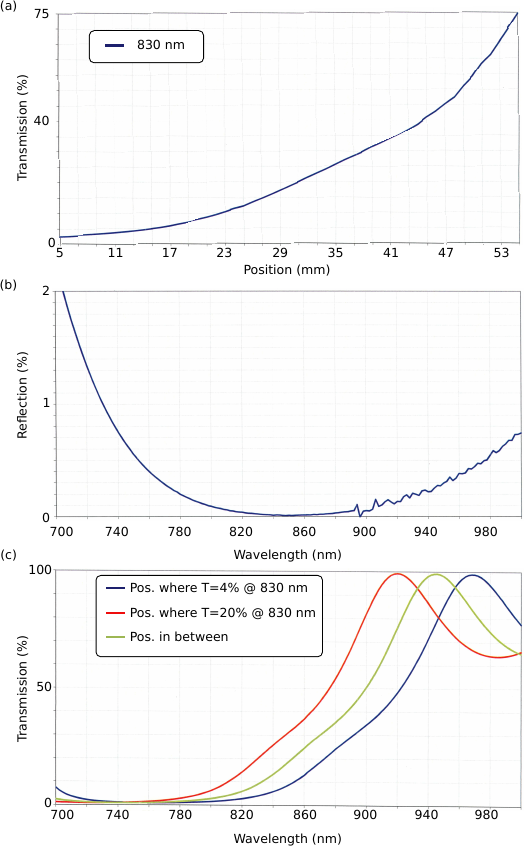}
    \caption{(a) Transmission versus vertical position at $830$~nm. The gradient coating increases transmission to normalize the intensity across emitters. (b) AR coating specifications for the input facet, which shows reflection of  $\sim 0.2\%$ at 780~nm. (c) Transmission at three vertical points versus wavelength shows substantially lower transmission at 780~nm, which enables the high resolutions seen in this work. Figure digitized from manufacturer testing results.}
    \label{sfig:coating}
\end{figure}

The VIPA has a gradient coating (Fig.~\ref{sfig:coating}), resulting in an approximately uniform intensity envelope across all emitters.
The output of the VIPA is expanded in the $y$ axis using a 4:1 cylindrical telescope. This magnification is chosen to tune the effective resolution of the grating, and produces an approximately square beam in the 29th order of the grating.

The resulting output has significant aberrations, primarily due to the many passes through the VIPA.
By far the largest aberration is associated with a slight wedge of the VIPA, which results in a departure angle for each emitter that scales linearly with the number of passes through the VIPA.
In the direction of the VIPA deflection, this results in a chirp in the positions of the emitters, and a slight nonlinearity that can be corrected with the appropriate conversion between frequency and position in the signal path. In the orthogonal direction, this results in astigmatism, which can be corrected by rotating one of the cylindrical lenses in the system.

The output of the grating is imaged onto an LCOS SLM via a 4f telescope with a demagnification of $1/4$, which allows one to compensate for the remaining aberrations in the system.
The SLM can additionally be used for amplitude modulation, to produce multiple copies of the image for parallelized operation across multiple logical qubits (e.g. for magic state distillation), and to scan the image produced by the system across a fiber for the fast video recording scheme described in Appendix~\ref{sec:singlepixel}.

In atom trapping applications, the output of the SLM could be imaged onto the atoms using a high NA microscope objective. Here, the output is imaged onto a camera for characterization. 

The overall efficiency of the system is 5.8\%, which is primarily limited by geometric constraints and the selected grating. An optimized design with a custom VIPA and grating could achieve efficiencies of $>50\%$.

\section{Image post-processing}
\label{sec:postprocessing}
To demonstrate the capabilities of our approach unconstrained by on-hand hardware, we apply several post-processing steps.
These are all equivalent to operations that can readily be performed optically in a future setup.
First, in Fig.~\ref{fig:setup}(c) we measure the optical transfer function by scanning the laser frequency and recording the PSF across the field of view. We then stitch together the PSFs with adjusted weighting to produce an arbitrary image.
Similarly, in Fig.~\ref{fig:hubbard}(a) we modulate individual FSRs and then compute the resulting image by summing the individual photos together.
These are both equivalent to modulating a wider RF bandwidth or wavelength division multiplexing.
Second, when modulating, we crop the carrier and negative sidebands from photos. This is equivalent to optically filtering or single sideband modulation with an extinction ratio surpassing our available MZM.
Finally, we compensate for ellipticity of the PSF by stretching the $y$ axis by $w_x / w_y$, based on a PSF measurement near the center of the field of view under the same conditions.
This is identical to a cylindrical telescope.
For each figure we fit a representative PSF under current conditions to determine the appropriate scaling factors.
An example fit can be seen in Fig.~\ref{fig:slm}(c).
These parameters are listed in Table \ref{tab:scaling} where $w_x$ and $w_y$ are the $\frac{1}{e^2}$ radii closest to the $x$ and  $y$ axes respectively, and $\theta_{\text{psf}}$ is the tilt of the PSF principal axes relative to the camera axes.
\begin{table}[htpb]
    \centering
    \caption{Measured PSF Parameters per Figure}
    \label{tab:scaling}
    \begin{tabular}{|c|c|c|c|}
        \hline
        Figure & $w_x$ &  $w_y$ & $\theta_{\text{psf}}$\\
        \hline
        \ref{fig:setup}(c),\ref{fig:setup}(d) & $\tileWx$ & $\tileWy$ & $\tileThetaDeg$\\
        \ref{fig:performance}(d) & $\chirpWxUm$ & $\chirpWyUm$ & $\chirpThetaDeg$\\
        \ref{fig:performance}(f) & $\interpWxUm$ & $\interpWyUm$ & $\interpThetaDeg$\\
        \ref{fig:performance}(g) & $\megabraidWxUm$ & $\megabraidWyUm$ & $\megabraidThetaDeg$\\
        \ref{fig:performance}(h) (y move) & $\chirpDefocusWxUm$ & $\chirpDefocusWyUm$ & $\chirpDefocusThetaDeg$\\
        \ref{fig:performance}(h) (x move) & $\interpDefocusWxUm$ & $\interpDefocusWyUm$ & $\interpDefocusThetaDeg$\\
        \ref{fig:hubbard}(a) (molecule) & $\harvardWxUm$ & $\harvardWyUm$ & $\harvardThetaDeg$\\ \ref{fig:hubbard}(a) (kagome, Lieb) & $\kagomeWxUm$ & $\kagomeWyUm$ & $\kagomeThetaDeg$\\ \ref{fig:slm}(a) & \psfWaistMajorUm & \psfWaistMinorUm & \psfThetaDeg\\
        \hline
    \end{tabular}
\end{table}

Note that these elliptical Gaussian fits, as well as those in Fig. \ref{fig:spatial_aberrations}, are performed by first fitting or estimating the PSF, then refitting restricted to a region two waists in radius around the predicted spot to focus on the central lobe. This slightly reduces the measured waist. 

\section{RF Control}
In this work we use two different signal paths optimized for different goals. Setup (A) uses a Tektronix AWG70001A AWG  operating at 48.8 GS/s with a \SI{20}{\giga\hertz} bandwidth and 7-bit vertical resolution. This is paired with two Lotus Systems LNA2G18G low noise amplifiers (LNAs). This setup operates between $2$~GHz and  $18$~GHz with a theoretical maximum output power of  $24$~dBm, and is used for characterizing large bandwidth operations such as moving spots and measuring modulation bandwidth, but struggles with linearity and vertical resolution. Vertical resolution is critical in our system for intensity control, since the finite resolution is shared across all tweezers. Likewise, crosstalk due to nonlinearity limits the effectiveness of our calibration procedure which assumes independence between tones.

Setup (B) uses a Xilinx ZCU216 RFSoC development board running the QICK firmware ~\cite{dingExperimentalAdvancesQICK2024}. This setup operates at 9.8 GS/s with 14-bit vertical resolution and is paired with a Mini-Circuits ZVE-3W-83+ $2$--$8$~GHz high-power linear amplifier and Mini-Circuits VHF-4600+ and VLF-8400+ filters to select the second Nyquist zone. This setup operates between $5.4$--$8$~GHz and has linearity limited only by the MZM itself.

\section{Balancing}
\begin{figure}[htbp]
    \centering
    \includegraphics[width=89mm]{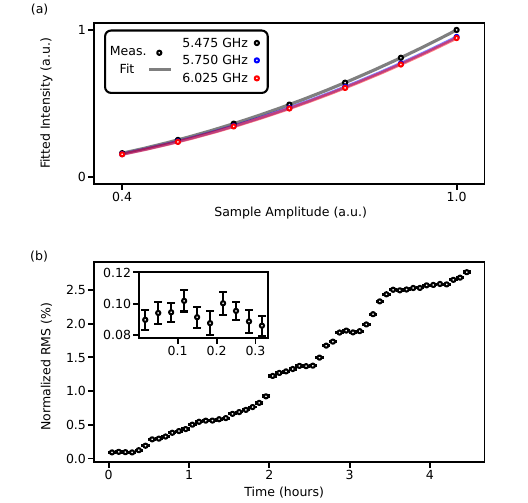}
    \caption{(a) Example balancing calibration curve with measured intensities and quadratic fit shown. (b) Normalized RMS of the balanced intensity versus time after calibration. Inset shows the first 20 minutes. Points are binned every 5 minutes for the main plot and every 2 minutes for the inset with mean and SEM shown.}
    \label{fig:balancing}
\end{figure}

To calibrate the mapping between RF power and optical intensity, we play a sum of tones corresponding to all targeted tweezers and measure the resulting intensity by fitting an elliptical Gaussian to the spot. It is helpful to crop the fit to an elliptical region around the predicted spot location, as in Appendix \ref{sec:postprocessing}, to minimize crosstalk. After measuring the mapping at several RF powers we fit a quadratic to the resulting curve, which is used for all subsequent amplitude modulation [Fig.~\ref{fig:balancing}(a)]. Since tweezers can be measured in parallel, it is possible to perform this calibration in $O(1)$ time for an arbitrarily large array. Calibrating once and computing the mean normalized RMS as a function of time reveals that without any particular effort to stabilize the system, the performance is maintained for at least $\sim 20$ minutes [Fig.~\ref{fig:balancing}(b)].

A remaining source of crosstalk in our current apparatus comes from the finite extinction ratio of the MZM and the presence of negative sidebands. The $\text{sinc}^2$ profile expected of the VIPA without apodization has power law tails which can be significant for tweezers that are closely spaced in the $y$ axis. For demonstration we ameliorate this issue by staggering the spots in the $y$ axis. However, this significantly limits the number of spots that can be balanced. Similar crosstalk also arises from the carrier, which is worsened by its increased intensity. We expect these issues to be preventing us from reaching error below $10^{-4}$, which would otherwise be within the vertical resolution of the RFSoC. Single sideband modulation as described in  Appendix~\ref{sec:postprocessing} and apodization of the PSF would resolve this issue.

\section{Iterative Aberration Correction}\label{sec:slm}
\begin{figure}[htbp]
    \centering
    \includegraphics[width=89mm]{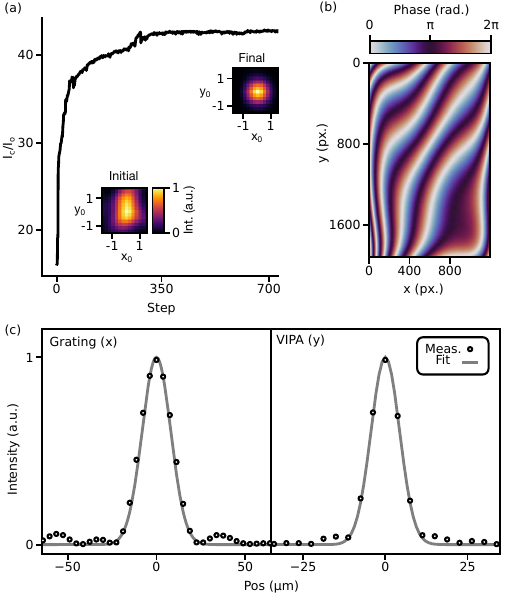}
    \caption{(a) Example SLM optimization curve of center-to-outer intensity ratio ($I_c / I_o$) versus step. Insets show initial and final PSF during optimization. (b) Resulting phase mask on the SLM. (c) Line cuts of the PSF after correction along the PSF major and minor axes. Measured values are computed by linear interpolation and plotted every pixel. Solid line indicates elliptical Gaussian fit.}
    \label{fig:slm}
\end{figure}

To estimate the optimal SLM correction, we apply a hill climbing procedure to each coefficient of a rectangular Legendre polynomial, up to total degree $5$, trying to maximize the ratio of the intensity within a $1$-pixel-radius region to that within a $20$-pixel-radius region. After each hill climbing iteration we fit a quadratic at the optimum which we accept if it improves the score, and repeat over these terms until convergence. This procedure converges quickly and is effective at recovering diffraction-limited performance. We include a representative optimization curve and resulting phase mask in Fig.~\ref{fig:slm}, along with line cuts of the 
resulting PSF. Future work may explore more sophisticated optimization techniques, including machine learning approaches or segment-based interferometric wavefront sensing \cite{zuoDeepLearningOptical2022,deutschNearfieldAmplitudePhase2008}.

\section{Dispersion Model and Spatial Resolution}
\begin{figure}[htbp]
    \centering
    \includegraphics[width=89mm]{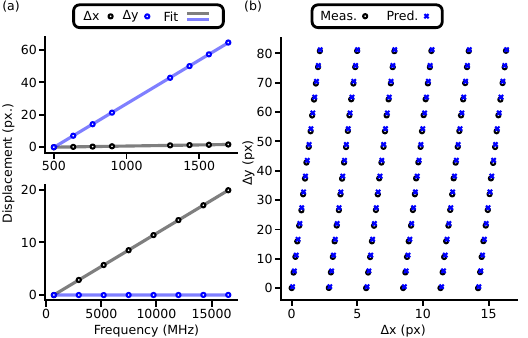}
    \caption{(a) Absolute displacement vs frequency within an FSR (top) and across FSRs (bottom). Linear fit shown. (b) Basic model predictions vs measured tweezer locations.}
    \label{fig:model}
\end{figure}

For generating waveforms and computing spatial resolutions we use a basic model of the system's dispersion, where we assume linear dispersion in both axes and orthogonality between the VIPA and the grating. To determine model parameters, we play a sequence of known frequencies from a DS Instruments SG22000PRO signal generator and fit to tweezer locations on the camera [Fig. \ref{fig:model}(a)]. By measuring the dispersion when moving within a single FSR and projecting the displacement onto the vector between two adjacent orders, we can estimate the position shift due to the VIPA alone. Based on the frequency needed to move between these orders we extract the FSR, and assuming the grating is orthogonal to the VIPA, we fit its dispersion across multiple FSRs. This procedure predicts tweezer locations within 1 camera pixel [Fig. \ref{fig:model}(b)], which is sufficient for our purposes, although it neglects nonlinearities in the VIPA and inexact orthogonality between the VIPA and the grating, so we expect systematic errors. Future megapixel-scale VIPA SLMs will likely demand more detailed, nonlinear models of the VIPA. Such models have been explored in spectroscopy literature \cite{xiaoDispersionLawVirtually2004} and can be directly applied to our system. 

We extract relevant parameters for resolution which we tabulate in Table \ref{tab:params}. We do not report uncertainties as shot noise in the camera is negligible and deviation is dominated by systematic errors in the model. These values together with the measured waists from Figs.~\ref{fig:setup}(c) and \ref{fig:setup}(d) in Table \ref{tab:scaling} are used to estimate the number of resolvable spots in the system. In particular, the frequency step $f_{\alpha}'$ which displaces by a waist is given by $ \frac{\delta \vec r_\alpha}{\delta f_\alpha} f_{\alpha}' = \vec w_{\alpha}^\parallel$ where $\vec r_\alpha$ is the displacement vector for the grating ($\alpha$ = g) or VIPA ($\alpha$ = v) and  $\vec w_\alpha^\parallel$ is the projection of the waist along the displacement vector (computed using $\theta_{\text{psf}}$ in Table \ref{tab:scaling}). Finally, the number of resolvable spots is given by $N_\text{v} = \left\lfloor \frac{\Delta f_\text{v}}{f_\text{v}'} \right\rfloor$ and $N_\text{g} = \left\lfloor \frac{\SI{10}{\tera\hertz}}{f_\text{g}'}\right\rfloor,\left\lfloor\frac{\tileSpanGHz}{f_\text{g}'}\right\rfloor,\left\lfloor\frac{\SI{40}{\giga\hertz}}{f_\text{g}'} \right\rfloor$ for the full C+L band, field-of-view limited span, and the 40 GHz span respectively. 

\begin{table}[htpb]
    \centering
    \caption{Fitted Dispersion Parameters}
    \label{tab:params}
    \begin{tabular}{|c|c|}
        \hline
        Param. & Value\\
        \hline
        $\Delta f_{\text{v}}$ & $\vipaFSRGHz$\\
        $\delta x/\delta f_{\text{v}}$ & $\vipaDispersionDxUmPerGHz$\\
        $\delta x/\delta f_{\text{g}}$ & $\echelleDispersionDxUmPerGHz$\\
        $\delta y/\delta f_{\text{v}}$ & $\vipaDispersionDyUmPerGHz$\\
        $\delta y/\delta f_{\text{g}}$ & $\echelleDispersionDyUmPerGHz$\\
        \hline
    \end{tabular}
\end{table}

\section{Waist across the field of view}
\begin{figure}[htbp]
    \centering
    \includegraphics[width=89mm]{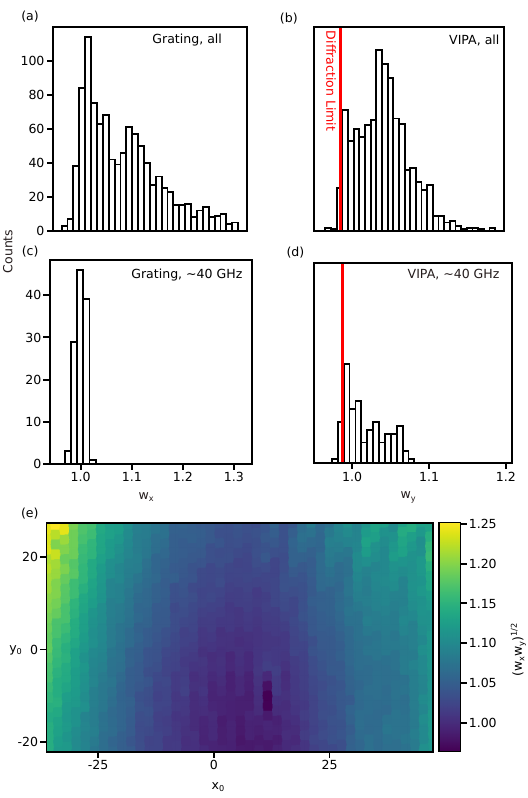}
    \caption{(a),(b) Histograms of waists along the grating and VIPA axes, respectively, normalized to the central reference waist in Fig \ref{fig:setup}(d). Red vertical lines on the VIPA histograms indicate diffraction limit. (c),(d) Same as (a),(b) but restricted to the central 17 FSRs accessible via the $\SI{40}{\giga\hertz}$ modulator. (e) Spatial map of the geometric mean of waists $\sqrt{w_x w_y} $ across the field of view with nearest neighbor interpolation.}
    \label{fig:spatial_aberrations}
\end{figure}

Fitting elliptical Gaussians to the spots shown in Fig.~\ref{fig:setup}(c), we find minimal variation in the waist across the field of view [Figs. \ref{fig:spatial_aberrations}(a)--\ref{fig:spatial_aberrations}(e)]. Note that these waists are measured with a static SLM correction near the center of the field of view. Some broadening is observed near the edges due to clipping of the beam on the 2-inch optics used in this work.

\section{Optimizing Movement} \label{sec:handoff}
To create a pure translation of a tweezer in the horizontal direction, we compensate for any tilt in the grating which would otherwise cause a change in vertical position when naively moving from $f_0$ to  $f_0 + \Delta f_{\mathrm{vipa}}$ by adding,
$$\text{Corr.} = - \frac{\delta y / \delta f_{\mathrm{g}}}{\delta y / \delta f_{\mathrm{g}} + \delta y / \delta f_{\mathrm{v}}} \Delta f_{\mathrm{vipa}} \ll \SI{1}{\mega\hertz}$$

Next, when doing handoffs across FSRs there is always broadening because the resulting spot is an interpolation of spatially separated tweezers. In this work, up to corrections for amplitude rolloff of the RF path with frequency, we use the functional form $V_1^2 = 1-s $ and $V_2^2 = s$ with $s \in [0,1]$ for the two tones to linearly move the center of mass. A downside to this approach is that it results in a non-constant waist during the move. In future work, with three overlapping tweezers, one can maintain constant broadening throughout (defined based on the second central moment) while linearly moving the center of mass. To achieve this with a triplet of adjacent spots $(V_{k-1}, V_k, V_{k+1})$ we can parametrize based on the offset of the center of mass $\Delta = x_{\mathrm{COM}} - x_k \in [-d/2, d/2]$ and the distance between adjacent spots $d$,
\begin{align*}
    V^2_{c\pm 1} &= \frac{1}{8} \left(1 + 4 \frac{\Delta^2}{d^2}\right) \pm \frac{\Delta}{2d}\\
    V^2_c &= \frac{3}{4} - \frac{\Delta^2}{d^2}
.\end{align*}
This gives linear motion of the COM and a constant waist of $w = \sqrt{w_0^2 + d^2}$ during the move. At the start/end of the move a two spot handoff can be used. Note that this is why there is no fundamental broadening when chirping within an FSR as $d\approx 0$.

This result implies a tradeoff between the bandwidth needed to cover a given distance along the grating axis and the resulting waist size during the move. However, this scaling is favorable in the sense that we must cover $N_w = \frac{w_0}{d}$ FSRs to move by one waist, so for small $\frac{d}{w_0}$ we get broadening,
\begin{align*}
    \frac{w}{w_0} &= \sqrt{1 + \frac{d^2}{w_0^2}} \approx 1 + \frac{1}{2} \frac{1}{N_w^2}  + O(N_w^{-4})
.\end{align*}
So broadening reduces quadratically with bandwidth devoted to the handoff.

\section{Adaptive Single-Pixel Camera}
\label{sec:singlepixel}
To record the dynamics of the dSLM we use the LCOS SLM already present in our system to scan the image plane across a $0.22$ NA bare fiber, whose output is focused onto a DC-400~MHz Thorlabs APD430A.
This is referred to as a single-pixel camera ~\cite{gibsonSinglepixelImaging122020}.
We average each trace over $1000$ experiments per pixel to determine a mean and uncertainty for the intensity at each timestep. 

With the optically conjugate CCD camera we reference the fiber location to a pixel by maximizing fiber coupling of a single spot and recording its location on the camera, doing a Gaussian fit for subpixel accuracy.
With this information, we can compensate for drifts in beam pointing and verify the point we are imaging is aligned within $0.2$px to the fiber location at each pixel we record. 

Traces are captured on a \SI{2}{\giga\hertz} LeCroy WaveRunner 204 MXi-A with a  $25$ sample linear phase finite impulse response (FIR) filter of $290$ MHz bandwidth.
In total the detector chain has a net bandwidth of $((\SI{290}{\mega\hertz})^{-2} + (\SI{400}{\mega\hertz})^{-2} + (\SI{2}{\giga\hertz})^{-2})^{-1 /2} \simeq \SI{233}{\mega \hertz}$. This implies a detector rise time of $\SI{1.5}{\nano\second}$, meaning we overestimate the intrinsic dSLM rise time by only $\simeq 0.1$~ns. 

\section{Modulation Bandwidth Characterization}
Although the frame rate is directly defined by the rise time, a related question is what spectral content can be transmitted through the system, for example to modulate links for the Hubbard scheme in Fig.~\ref{fig:hubbard}. In the dSLM, we can consider a sinusoidal amplitude modulation in the steady state as the sum of a carrier and two sidebands. As the frequency increases, the two sidebands are increasingly dispersed, effectively attenuating the modulation. 

\begin{figure}[htbp]
    \centering
    \includegraphics[width=89mm]{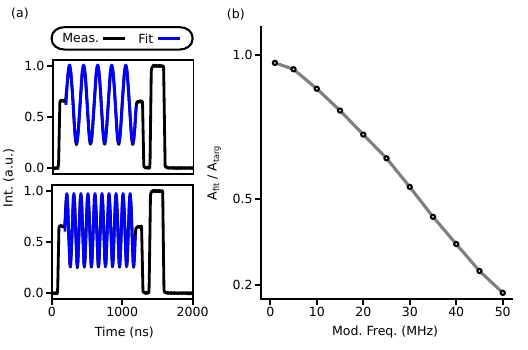}
    \caption{(a) Example traces at $5$~MHz (top) and  $10$~MHz (bottom) captured on the single-pixel camera. Fitted amplitude is plotted in blue. (b) Fit amplitude $A_{\mathrm{fit}}$ normalized by target amplitude  $A_{\mathrm{targ}}=0.4$ as a function of modulation frequency.}
    \label{fig:bw}
\end{figure}

To quantify this, we apply a sinusoidal modulation with a normalized amplitude from $0.2$ to $1.0$ on top of a carrier frequency of $9.5$~GHz for the maximum number of periods that can fit in $1.01$ $\mu$s. For each modulation frequency, this is followed by a pulse to amplitude $1.0$ for normalization, which is also used for the rise time measurement. We record near the center of the resulting spot on our single-pixel camera and accumulate traces at frequencies between $1$~MHz and  $50$~MHz, and observe the rolloff in amplitude with increasing frequency (Fig. \ref{fig:bw}).

\section{Dynamics beyond the bandwidth limit}
\label{sec:phasejump}
\begin{figure}[htbp]
    \centering
    \includegraphics[width=89mm]{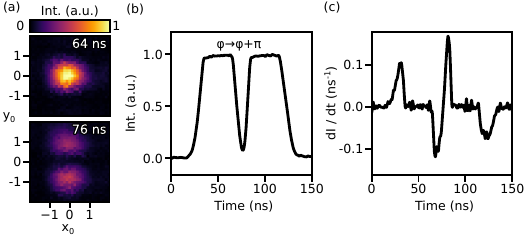}
    \caption{(a) Selected frames in steady state (top) and after a $\pi$ phase jump (bottom).  The spatial profile after the phase jump is consistent with the interference of two spots with halved resolution as expected. (b) Intensity summed along $y_0 = 0$, normalized to $[0,1]$. A sudden extinction is seen after the phase jump, followed by a revival to steady state. (c) Numerical derivative of the intensity in (b). Comparing phase jump to the regular rise/fall, we see the maximum derivative when turning on to be $\sim\phaseJumpPeakRatio \times $ larger and for turning off to be $\sim\phaseJumpTroughRatio \times $ larger.}
    \label{fig:phasejump}
\end{figure}

Although the response time of the system is set by the time for light to traverse the VIPA, dynamics beyond this limit can be observed by considering the transient response of light propagating through the system. An example of this is shown in Fig.~\ref{fig:phasejump}, where we apply a $\pi$ phase jump to a tweezer and observe the time evolution in the image plane. At the time when half of the light with the phase jump has traversed the VIPA, the resulting pattern will be the interference of two spots with halved resolution, producing a dark region at the center of the original tweezer.
This is desirable when one wants to momentarily extinguish a tweezer faster than would be possible with the regular bandwidth, for example to avoid light shifts during a gate operation.
More sophisticated control of the phase and amplitude of the light in the VIPA can allow for more complex dynamic operations in future work.

\section{Optical simulations and parameter optimization}

Beam propagation through the VIPA is simulated using angular-spectrum propagation.
To identify the appropriate incoupling beam parameters (namely beam waist, position, and angle) a disorder-free 1D simulation is used, allowing for optimization via gradient descent or brute force search.
The high reflection cutoff on the input facet is modeled as a hyperbolic tangent function in reflectivity with a transition edge of $2~\mu m$, as is representative of high quality commercial VIPAs.
To certify performance in the presence of disorder, a 2D simulation is used, where surface roughness is modeled as a spatially correlated displacement map on each surface.
To model a static correction of the resulting phase errors with an LCOS SLM, we sample the phase errors at a single reference frequency, and apply a correction that is discretized to a $4160\times2464$ pixel grid (representative of the highest resolution LCOS SLMs that are readily available).
Although the SLM pattern remains fixed, as the laser frequency is varied both the phase error from the VIPA and the correction applied by the SLM shift slightly.
For very large frequency changes, the correction applied by the SLM will no longer be appropriate.

\section{Numerical method of computing the Hubbard parameters}
\label{sec:mlwf} 

To simulate tweezer arrays generated by the dSLM we assume that each tweezer has a Gaussian beam profile of equal width $w_0$. Position $\mathbf{r}_{i}$ and depth $D_{i}$ are free parameters that are controlled by the RF drive. 
Because neighboring tweezers are well separated in frequency in the dSLM setup, we assume an incoherent sum of intensities of the tweezer traps. The total potential is therefore: 
\begin{align*}
	V_{total}(\mathbf{r}) = \sum_{i} -D_{i}\exp{(-2(\mathbf{ r} - \mathbf{r}_{i})^2/w_0^2)}
\end{align*}

\begin{figure}[htbp]
 	\centering
 	\includegraphics[width=89mm]{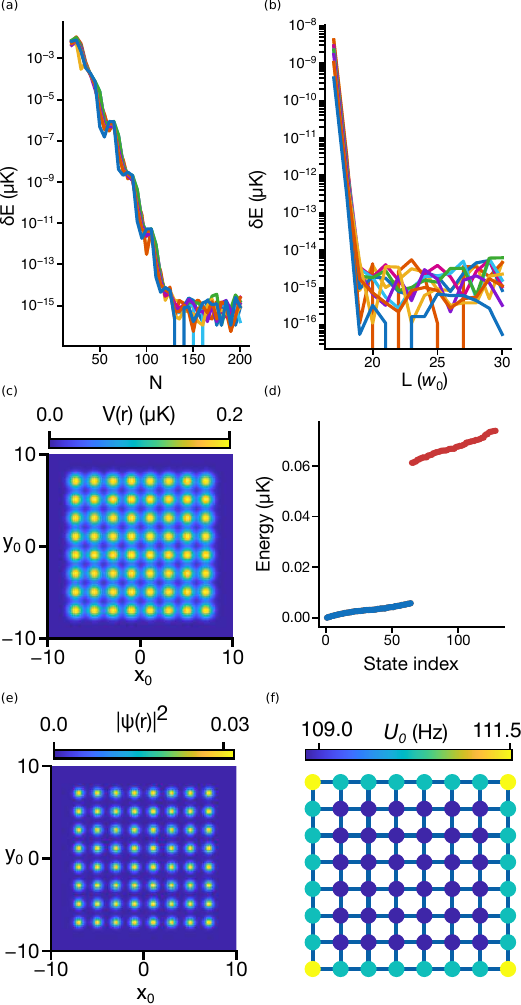}
 	\caption{Numerical method of computing the Hubbard parameters. (a) Convergence of DVR energy with number of grid points $N$ for a 15-by-15 square lattice with edge length $L=20w_0$. The first 10 eigenenergies are shown as an example. (b) Convergence of DVR energy with system size $L$ for a 15-by-15 square lattice. Grid spacing is fixed at $\Delta x = 2/15 w_0$. (c) Simulated optical potentials of an 8-by-8 programmable square lattice. The 8-by-8 base tweezers have a depth of 0.2~$\mu$K and the $t$-control tweezers have a depth of 0.04~$\mu$K. (d) Energy spectrum in the DVR basis for the first 128 states. In all calculations, we keep the depth of the control tweezers small enough to avoid band mixing. The number of states in the lowest band (blue) matches the number of base tweezers.  We truncate the maximally localized orbitals calculation to the states in the lowest band. (e) Maximally localized single particle orbitals from the DVR basis in the lowest band. All 64 states are well localized on the base tweezers. (f) Computed Hubbard parameters from the single particle orbitals. Bond thickness represents the tunneling strength $|t|$ and fill colors on sites represent the on-site interaction strength $U_0$.}
 	\label{fig:MLWF}
 \end{figure}

 We adapted the discrete variable representation (DVR) method used in Refs.~\cite{wallEffectiveManybodyParameters2015, weiHubbardParametersProgrammable2024} to construct the eigenbasis of the tweezer array. We assume a 2D tweezer potential with fixed confinement in the $z$ direction across all sites. The DVR discretizes the 2D plane and solves the Schr\"{o}dinger equation given the optical potential. Figs.~\ref{fig:MLWF}(a) and \ref{fig:MLWF}(b) show a convergence test in a square lattice geometry of 15-by-15 tweezers.
 Unless otherwise stated, we use a square system with edge length $L = 20w_0$, sampled on a grid with $N_x=N_y=150$ points.

 For most of the geometries, we truncate the number of eigenstates to the number of tweezer traps to focus on only the lowest band.
 Figs.~\ref{fig:MLWF}(c) and \ref{fig:MLWF}(d) show an example of the potential and energy spectrum of eigenstates in the DVR basis.
 The lowest band, spanned by the number of tweezers, is well-separated from the higher bands.
 Using the single-particle eigenstates of the optical potential for a finite tweezer array, we construct a set of maximally localized single-particle orbitals (Wannier-like functions) associated with individual tweezers.
 Starting from an orthonormal basis $\{\phi_k(\mathbf r)\}_{k=1}^{M}$ spanning the relevant low-energy subspace, we generate a new orthonormal set $\{W_n(\mathbf r)\}$ via a unitary rotation,
\begin{align*}
 	W_n(\mathbf r)=\sum_{k=1}^{M}\textbf{U}_{nk}\,\phi_k(\mathbf r), \qquad \textbf{U}\in \textbf{U}(M).
 \end{align*}
 We determine $\textbf{U}$ using the Foster--Boys localization criterion, which chooses the rotation to minimize the spatial extent of the orbitals. Defining the center of each orbital as
\begin{align*}
 	\langle x\rangle_n=\langle W_n|x|W_n\rangle,\;\;\langle y\rangle_n=\langle W_n|y|W_n\rangle,
 \end{align*}
 we maximize $F$, which is equivalent to minimizing the total quadratic spread:
\begin{align*}
 	F=\sum_{n}\left(\langle x\rangle_n^{2}+\langle y\rangle_n^{2}\right).
  \end{align*}
 Finally, we assign each localized orbital $W_n(\mathbf r)$ to a specific tweezer site by matching its center $\mathbf r_n$ to the nearest tweezer position. We calculate the Hubbard parameters, including tunneling strength $t_{ij}$, on-site interaction $U_{0}$, and chemical potential $\mu$, by integrating over the real space coordinates. Figs.~\ref{fig:MLWF}(e) and \ref{fig:MLWF}(f) show an example of the computed maximally localized single-particle orbitals and Hubbard parameters.
 
 \section{Hubbard parameter tuning and optimization}
\label{sec:hubbard_opt} 
 
 \begin{figure*}[!htb]
 	\centering
 	\includegraphics[width=180mm]{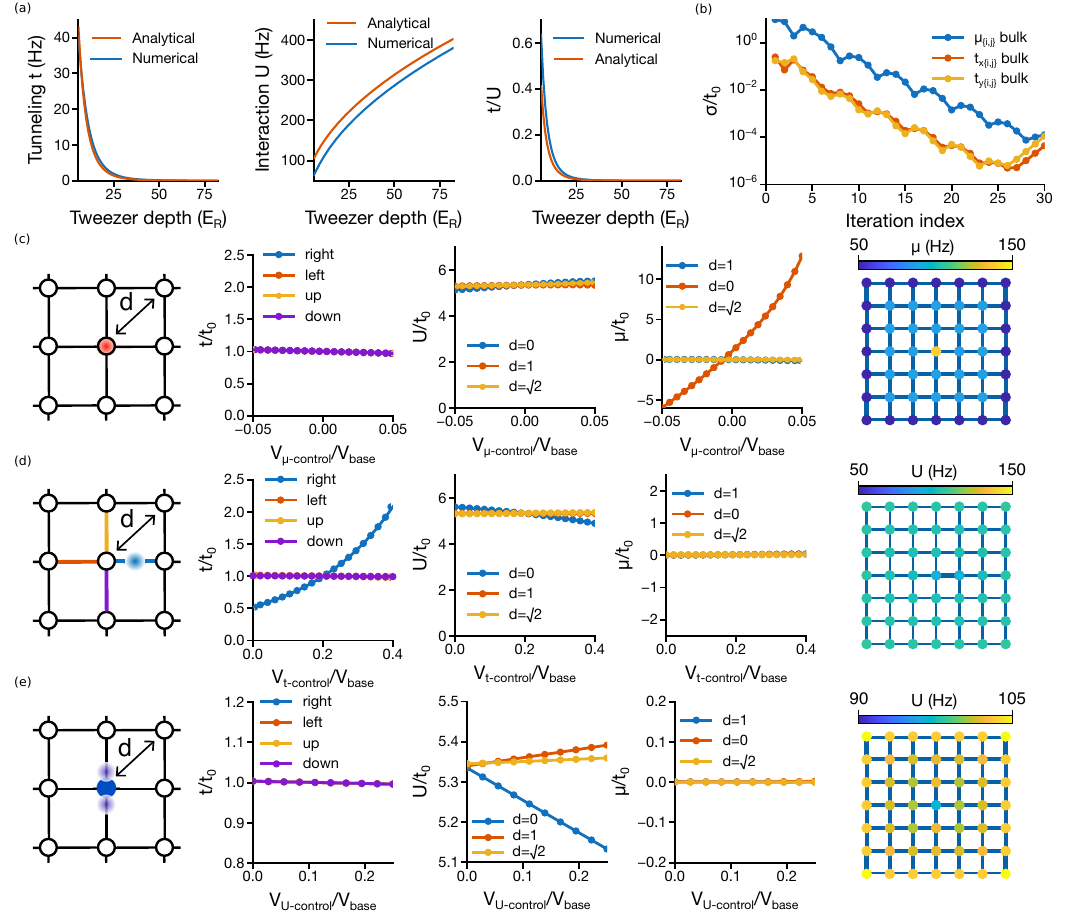}
 	\caption{Numerical simulation of programmable Hubbard parameters in a square lattice. (a) Comparison of the computed Hubbard parameters in the tweezer array to the analytical solution in a sinusoidal square lattice shows good agreement. (b) Standard deviation of the Hubbard parameters in the bulk as a function of iteration step of the feed-forward algorithm. Iteration step = 20 is used for optimizing Hubbard parameters in the example shown in the main text. (c)-(e) Tuning parameters on the central site in a 7-by-7-site system. Left to right: cartoon indicating the convention followed by the labels in the legend; dependence of tunneling strength, interaction strength, and chemical potential on control tweezer depth; characteristic example of all Hubbard parameters of the system at a given setting. (c) shows chemical potential tuning, (d) tunneling strength, and (e) interaction strength.
    }
 	\label{fig:hubbard_tuning}
 \end{figure*}
 
Using the DVR method, we construct a basis of maximally localized single-particle orbitals and evaluate the corresponding Hubbard parameters. Throughout this work, we use \(^{87}\mathrm{Rb}\) as a representative atomic species. We assume a diffraction-limited optical system with center wavelength \(\lambda = 1064\,\mathrm{nm}\) and numerical aperture of \(\mathrm{NA}=0.55\), giving a diffraction-limited beam waist:
\begin{align*}
	w_0 = \frac{\lambda}{\pi \mathrm{NA}} .
\end{align*}

The simulated dSLM-projected potentials are restricted to the transverse \(x\)-\(y\) plane. Confinement in the out-of-plane direction is assumed to be harmonic, with a trap frequency of \(\omega_z = 2\pi \times 30\,\mathrm{kHz}\).
The finite spatial extent of the simulated lattice leads to appreciable boundary effects: edge sites generally acquire large local energy offsets relative to bulk sites. To avoid contamination of the extracted bulk parameters, we exclude Hubbard parameters associated with the outermost sites from all subsequent analysis. These boundary sites are retained in the numerical calculation as ``ghost'' sites, which suppress finite-size artifacts and improve convergence of the bulk Wannier-like orbitals and Hubbard matrix elements~\cite{weiHubbardParametersProgrammable2024}. Figure~\ref{fig:hubbard_tuning}(a) compares the DVR-extracted nearest-neighbor tunneling matrix element and on-site interaction energy as functions of the base tweezer depth with the corresponding analytical expressions for a simple interfering optical lattice~\cite{blochManybodyPhysicsUltracold2008}.

For the programmable square lattice, we supplement the base tweezer array with two classes of auxiliary control tweezers: $t$-control tweezers and $U$-control tweezers. These auxiliary tweezers introduce controlled perturbations to the base potential and enable local modification of the effective Hubbard parameters:

\paragraph{Chemical potential.}
The local chemical potential is set by the single-particle ground-state energy of each trap, including both the harmonic zero-point contribution and the local potential offset. In the perturbative regime, this parameter is controlled predominantly by the depth of the corresponding base tweezer.

\paragraph{Tunneling.}
The nearest-neighbor tunneling amplitude is determined primarily by the inter-site separation and by the height and curvature of the potential barrier between adjacent base traps. We use $t$-control tweezers positioned between neighboring lattice sites to locally modify the barrier height, thereby tuning the corresponding tunneling matrix element.

\paragraph{On-site interaction.}
The on-site interaction energy is determined by the spatial extent of the localized orbital and therefore depends on the local trap curvature. Independent control of \(U\) requires modifying the orbital confinement while minimizing changes to the local potential minimum and the inter-site barriers that determine \(t\). To this end, we introduce $U$-control tweezers at non-integer spacing relative to the base and $t$-control tweezer positions. Since the dSLM provides continuous displacement only along the VIPA-dispersed axis, the $U$-control tweezers are placed at a displacement of \(0.6w_0\) from the corresponding base tweezer along this direction. This spacing maximizes the available single-site dynamic range for tuning \(U_0\) within the allowed geometry.
Greater tuning range is achievable at the cost of dSLM resolution by placing multiple rows within a beam waist in the grating dispersed direction.
In this case a larger dynamic range can be obtained by placing four $U$-control tweezers around each base site, for example near the SW, NW, SE, and NE quadrants, which allows for more symmetric control of the local curvature while suppressing linear shifts of the trap center.
Additional approaches, including site-resolved control of collisional properties~\cite{pampelQuantifyingLightAssistedCollisions2025} and time-periodic modulation of the local potential~\cite{cardarelliEngineeringInteractionsAnyon2016}, could also be integrated with the dSLM architecture to extend the accessible range of local \(U_0\) tuning.

In practice, the above controls are coupled.
For example, varying the depth of any tweezer changes the local curvature of the trapping potential and therefore the spatial extent of the associated localized orbital.
The same perturbation also shifts the local chemical potential and modifies nearby tunneling energies.
These undesired changes can be compensated perturbatively through corrections to the base tweezer depths and $t$-control tweezer amplitudes. We use an algorithm that computes the Hubbard parameters after each step, and feeds the difference from the target values to the next step [see Fig.~\ref{fig:hubbard_tuning}(b)]. 
Note that due to the smaller tuning range of $U$, and complicated couplings induced by adding $U$-control tweezers to all sites, we do not include $U$ compensation on all sites.
The tuning range of each parameter on a single site or link is shown in Figs.~\ref{fig:hubbard_tuning}(c)--\ref{fig:hubbard_tuning}(e).
 
\section{Artificial magnetic field with programmable local flux}
\label{sec:local_flux} 
 While most implementations of artificial magnetic fields modulate the local chemical potential, we are able to modulate the tunneling energy on each link to achieve the same effective Hamiltonian. This scheme allows for direct and independent access to a complex phase on each link.

One can use this modulation to realize an artificial magnetic field in a Landau gauge, which we show below. Note that the same derivation applies for the symmetric gauge. \\
The time-dependent Hamiltonian in the tight-binding model is given by:
 \begin{align*}
 	\begin{aligned}
 		\hat{H}(t) &= - \sum_{ij}t^{(x)}_{ij}(t)\hat{a}^{\dagger}_{i+1,j}\hat{a}_{i,j} + h.c. - \sum_{ij}t^{(y)}\hat{a}^{\dagger}_{i,j+1}\hat{a}_{i,j}  \\
 		& + h.c. + \frac{U}{2}\sum_{ij} \hat{n}_{i,j}(\hat{n}_{i,j}-1) + \sum_{ij}\Delta_i\hat{n}_{i,j}
 	\end{aligned}
\end{align*}
 We assume a sinusoidal drive of the $x$-links with frequency $\omega$, and a linear tilt in the $x$-direction that is either directly programmed using control over the local potential, or realized using a magnetic field gradient leading to a spatially varying Zeeman shift.
 \begin{align*}
 	t^{(x)}_{ij}(t) = t^{(x)} +\delta t_{ij}\cos(\omega t+\phi_{ij})
\end{align*}
 \begin{align*}
 	\Delta_i = i\Delta_{tilt}
\end{align*}
 The driving frequency is resonant with the tilt, i.e., $\hbar \omega = \Delta_{tilt}$. In the high-frequency limit, the driving frequency is the highest energy scale in the Hamiltonian: $\hbar\omega \gg t^{(x)},t^{(y)},U $. The time-dependent Hamiltonian in the rotating frame is: 
 
 \begin{align*}
 	\begin{aligned}
 		&\hat{H}_F(t) =  -\sum_{ij} [t^{(x)}e^{i\omega t} +\frac{\delta t_{ij}}{2}(e^{i2\omega t+i\phi_{ij}} + e^{-i\phi_{ij}})]\hat{a}^{\dagger}_{i+1,j}\hat{a}_{i,j}  \\
 		& + h.c. -\sum_{ij}t^{(y)}\hat{a}^{\dagger}_{i,j+1}\hat{a}_{i,j} + h.c. + \frac{U}{2}\sum_{ij} \hat{n}_{i,j}(\hat{n}_{i,j}-1)
 	\end{aligned}
\end{align*}
 The time-independent effective Hamiltonian is obtained by performing the Magnus expansion to lowest order: 
 
  \begin{align*}
 	\begin{aligned}
 		\hat{H}_{eff} &\simeq \frac{1}{T}\int \hat{H}_{F}(t)dt = -\sum_{ij} \frac{\delta t_{ij}}{2} e^{-i\phi_{ij}}\hat{a}^{\dagger}_{i+1,j}\hat{a}_{i,j} + h.c. \\
 		& -\sum_{ij}t^{(y)}\hat{a}^{\dagger}_{i,j+1}\hat{a}_{i,j} + h.c. + \frac{U}{2}\sum_{ij} \hat{n}_{i,j}(\hat{n}_{i,j}-1)
 	\end{aligned}
\end{align*}
 This effective Hamiltonian is the interacting Harper--Hofstadter model (HHM) in the Landau gauge when $\delta t=2t^{(y)}$.
 The flux $\Phi_{p,q}$ enclosed by the plaquette of (i,j), (i+1,j), (i, j+1), (i+1,j+1) is defined as: 
  \begin{align*}
 	\Phi_{p,q}/\Phi_0 = \frac{\phi_{i, j+1} - \phi_{i,j}}{2\pi}
\end{align*}
 $\Phi_0$ is a flux quantum. When $\Phi_{p,q}$ are identical for all plaquettes, we obtain a uniform artificial magnetic field. 
 
 In a conventional artificial magnetic field, the flux is set by the $\mathbf{k}$-vector of the light potential that generates the gauge field, i.e., $\phi_{ij} = \mathbf{k \cdot } \mathbf{r}_{ij}$. This leaves no flexibility in local degrees of freedom. In the dSLM setup, the phase on each link can be programmed independently, leading to a programmable flux on each plaquette. For example, we can insert a flux on only a single plaquette in the Landau gauge. Specifically, the HHM in the Landau gauge takes the form: 
 \begin{align*}
 	\hat{H}_{HH} = -\sum_{ij} t^{(x)} e^{-i\phi_{ij}}\hat{a}^{\dagger}_{i+1,j}\hat{a}_{i,j}  + h.c. \\ -  \sum_{ij}t^{(y)}\hat{a}^{\dagger}_{i,j+1}\hat{a}_{i,j} + h.c.
 \end{align*}
 To change the flux on a single plaquette, we can change the phase difference of the top and bottom links:  
 \begin{align*}
 	\Phi_{p,q}^\star/\Phi_0 &= \frac{(\phi_{i,j+1}+\delta\phi) - \phi_{i,j}}{2\pi} \\
 	& = \Phi_{p,q}/\Phi_0 + \delta\phi/2\pi
 \end{align*}
However, this leads to a change of flux on the neighboring plaquette with opposite sign: 
 \begin{align*}
 	\Phi_{p,q+1}^\star/\Phi_0 &= \frac{(\phi_{i,j+2} - (\phi_{i,j+1}+\delta\phi))}{2\pi} \\ &=\Phi_{p,q+1}/\Phi_0  - \delta\phi/2\pi 
 \end{align*}
 To maintain a uniform flux background, we can apply a $\delta\phi$ change to all other links along this line, extending to the boundary of the system.
 Note that the topology of a localized flux is precisely reflected by the fact that we needed to introduce a defect extending to the boundary of the system in order to generate it.

\section{AI Usage}
GitHub Copilot,  Google Gemini 3.0 Pro and 3.1 Pro, and Anthropic Claude Sonnet 4.6 assisted in developing some subroutines in the control and measurement code. Copilot and Gemini were additionally used in developing portions of the angular-spectrum propagation code and scripts that loop over simulation runs. OpenAI ChatGPT 5.5 Thinking assisted with conceptualizing ideas, writing subroutines, and debugging in the Hubbard model simulation. Claude Fable 5, Opus 4.8, and ChatGPT 5.5 Thinking assisted in developing plotting code. All AI tools were used interactively, with the authors continuously validating the output.
\end{document}